\documentclass[a4paper,11pt]{article}
\usepackage{jcappub} 
\pdfoutput=1 
\usepackage[T1]{fontenc} 
\usepackage{lineno,ulem, caption,subcaption,multirow,physics,siunitx}
\newcommand{\Slash}[1]{{\ooalign{\hfil/\hfil\crcr$#1$}}} 

\arxivnumber{2412.03219} 

\title{\boldmath Neutron star in Logarithmic model of Cartan $F(R)$ gravity}

\author[1]{Hiroki Sakamoto}
\affiliation[1]{Research Institute for Semiconductor Engineering, Hiroshima University,\\
1-4-2 Kagamiyama, Higashi-Hiroshima 739-8527 JAPAN}
\author[2]{Masahiko Taniguchi}
\affiliation[2]{Graduate School of Advanced Science and Engineering, Hiroshima University, \\
Higashi-Hiroshima, 739-8526, Japan}

\emailAdd{h-sakamoto@hiroshima-u.ac.jp}
\emailAdd{masa-taniguchi@hiroshima-u.ac.jp}

\abstract{Cartan $F(R)$ gravity introduces the equivalent scalar-tensor theory by extending the gravity sector.
From the solution of the modified Cartan equation leads to the interaction with the scalar field and the fermion.
We derived the effective potential of the scalar field using the auxiliary field method, which is commonly used in studies of spontaneous chiral symmetry breaking.
Using this effective potential we investigated the mass-radius relation of neutron star to solve the Tolman-Oppenheimer-Volkoff equation.
By the numerical computation, we found that the contribution of the scalar field increases the mass of a neutron star.
This could give theoretical plausibility to the current observations.}

\begin{document}
\maketitle
\flushbottom

\section{Introduction}
The success of Einstein's general relativity (GR) is confirmed by such as the observation of gravitational waves, however, the gravitational theories modified GR are still under consideration.
The modification of GR is required to explain the phenomena in the strong field region such as inflation, black holes, or neutron stars.
One of the simplest modifications of modified gravity is known as $F(R_E)$ gravity, in which the Lagrangian of this model contains the arbitrary function of Ricci scalar $R_E$ ~\cite{Clifton:2005aj,Nojiri:2006ri,Nojiri:2010wj,Nojiri:2017ncd}.
The inclusion of the $R_E^2$ term to Einstein-Hilbert action, which is called Starobinsky model, can explain the fluctuation observed from the cosmic microwave background (CMB) ~\cite{Starobinsky:1980te}.
The fundamental dynamical variable of $F(R_E)$ gravity is the metric.
The variational principle derives from the Einstein equation.
Models constructed by the scalar curvature are called the metric formalism.
While the Cartan formalism, which is formulated by localization of the Lorentz transformation,
assumes the two dynamical variables, i.e., the vierbein and the spin connection
~\cite{Kibble1961jmp,SCIAMA:1964rmp,Poplawski:2010jv,Magueijo:2012ug}.
The variational principle w.r.t. spin connection leads to the Cartan equation.
However, this equation is non-dynamical in the case of Einstein-Hilbert action, thus metric formalism is reproduced.
The extension of Einstein-Hilbert action to the arbitrary function of scalar curvature in Cartan formalism leads to the fact that the Cartan equation has dynamics. This model is called Cartan $F(R)$ gravity.

In the metric $F(R)$ gravity, the gravitational sector can be reduced to the Einstein-Hilbert term
by the conformal transformation.
A new DoF which is yielded by the modification can be represented by the scalar field.
Applying the conformal transformation to the Cartan $F(R)$ gravity, we can also obtain the Einstein-Hilbert term
and the scalar field.
However, the scalar field is not the same as the one in the metric $F(R)$ gravity
since the corresponding theory is the Brans-Dicke theory with $\omega = -3/2$
and, in this theory, the scalar field carries no dynamics of its own~\cite{Sotiriou:2008rp}.
From this, we can not obtain the equivalent scalar-tensor theory
by applying the conformal transformation to the Cartan $F(R)$ gravity.
However, instead of the conformal transformation,
we can reduce the model to the Einstein-Hilbert term of metric formalism and a scalar field using the variational principle of spin connection in Cartan $F(R)$ gravity.
This scalar field has interaction with the spinor through the four-fermion interaction.
The four-fermion interaction appears in the model can be deformed to the
Nambu-Jona-Lasinio (NJL) model by the Fiertz transformation.
To incorporate the contribution of the spinor field, it is easy to consider
the effective potential, which is derived by integrating out the spinor field.
In the case that the vacuum expectation value (VEV) of $\bar\psi\psi$ has a non-zero value,
the correction term can be obtained.

In this paper, we investigate the contribution of the scalar field induced by the Cartan $F(R)$ gravity on neutron stars.
The study of neutron stars is essential across nuclear, particle, and gravity physics. 
Their extreme density and curvature provide various indications to explore dense matter equations of state, exotic particles, and deviations from GR. 
In particular, neutron stars provide an important testing ground for modified gravity theories in strong gravitational fields. 
Previous studies have been carried out in metric $F(R)$ gravity. gravity~ \cite{Astashenok:2013vza,Astashenok:2014nua,Astashenok:2017dpo,Astashenok:2020cfv,Astashenok:2020cqq,Astashenok:2021peo,Astashenok:2021xpm,Astashenok:2021btj,Capozziello:2015yza,Odintsov:2023ypt,Numajiri:2021nsc,Cui:2024nkr}.
Recent neutron observations have been advanced by gravitational waves~\cite{LIGOScientific:2017vwq} and pulsars~\cite {Miller:2019cac}.
Various observations have not reported the existence of neutron stars below $1M_{\odot}$~\cite{Lattimer:2012nd}.
We have obtained results that increase the minimum mass of neutron stars by the scalar field of Cartan $F(R)$ gravity theory.
This study provides a plausible explanation for the observation results.

This paper is organized as follows.
In Sec.2 we introduce the Cartan $F(R)$ gravity with a spinor field.
Using the modified Cartan equation and integration out for the spinor field,
we derive the effective potential of the scalar field.
In Sec.3, we consider the Tolman–Oppenheimer–Volkoff (TOV) equation with the scalar field.
In Sec.4, we shall show the numerical results of the TOV equation in some parameters.
The mass-radius relations are figured and we can see that
the effect of the scalar field makes the mass of the neutron star heavy.
Finally, we give some concluding remarks and discussion.

\section{Cartan $F(R)$ gravity with a spinor field}
The gauge symmetry of elementary particles is the fundamental concept of building the
a new model of modern physics, the localization of Lorentz transformation is a natural way to
consider as the new physics.
The spinor field $\psi$ is the matter field which has the $SL(2, \mathbb{C})$ group as a representation of the Lorentz transformation.
The generator of this group is given by the gamma matrix $\gamma^i$, which satisfies the Clifford algebra $\{\gamma^i, \gamma^j\} = 2\eta^{ij}$, where $\eta^{ij}$ is the Minkowski metric.
Hereafter the alphabet index denotes the flat Minkowski space-time, and the Greek index denotes the curved space-time.
The covariant derivative of the spinor field is defined by,
\begin{align}
   \label{eqs::cov dev spinor}
  D_\mu \psi = \partial_{\mu} \psi + \frac{1}{4} \omega_{ij\mu} \gamma^i \gamma^j \psi,
\end{align}
where $\omega_{ij\mu} = -\omega_{ji\mu}$ is the spin connection.
The local Lorentz invariant Lagrangian for the massless spinor field is given by,
\begin{align}
   \label{eqs::Lagrangian::spinor}
  \mathcal{L}_\psi = -\frac{i}{2} \bar{\psi} \Big[ \Slash{D} - \overleftarrow{\Slash{D}} \Big] \psi.
\end{align}
The field strength of the spin connection is defined by the curvature tensor $R_{\mu\nu}^{ij}$ as follows,
\begin{align}
  \label{eqs::curvature tensor}
  {R^{ij}}_{\mu\nu} = \partial_{\mu}{\omega^{ij}}_\nu-\partial_{\nu}{\omega^{ij}}_\mu+{\omega^i}_{k\mu}{\omega^{kj}}_\nu-{\omega^i}_{k\nu}{\omega^{kj}}_\mu.
\end{align}
The flat Minkowski space-time is tangent to the curved space-time at each point.
The vierbein ${e^i}_\mu$ is the basis of the tangent space, which connects the curved metric $g_{\mu\nu}$ and the flat metric $\eta_{ij}$ as,
$g_{\mu\nu} = \eta_{ij} {e^i}_\mu{e^j}_\nu$.
%
%
To contract the \eqref{eqs::curvature tensor} with the vierbein, we obtain the curvature scalar $R$ as follows,
\begin{align}
  \label{eqs:def of Ricci tensor}
  R = {e_i}^\mu{e_j}^\nu {R^{ij}}_{\mu\nu}(\omega, \partial\omega).
\end{align}
The minimal model, which is included only \eqref{eqs::Lagrangian::spinor} and \eqref{eqs:def of Ricci tensor} in the action,
is called the Einstein-Cartan-Sciama-Kibble (ECSK) theory and
It is the extended model of the GR with spinor field which means the Riemann curvature tensor is replaced by the curvature scalar with the spin connection.
However, the ECSK theory is equivalent to the GR in the sense of the gravity theory.
The curvature scalar can be decomposed into the Riemann curvature scalar $R_E$ and other parts.
We introduce the affine connection $\Gamma^\rho_{\mu\nu} = {e_a}^\rho D_\nu {e^a}_\mu$
and the torsion tensor $T^\rho_{\mu\nu} = \Gamma^\rho_{\mu\nu} - \Gamma^\rho_{\nu\mu}$.
The curvature scalar can be expressed by the $R_E$ and the torsion part as,
\begin{align}
   \label{eqs:Ricci scalar into non and torsion2}
  R = R_E + T - 2{\nabla_E}_\mu T^\mu,
\end{align}
where $\nabla_E$ is the covariant derivative expressed by the Levi-Civita connection.
$T_\mu$ means the torsion vector, $T_\mu={T^\lambda}_{\mu\lambda}$, and the torsion scalar $T$ is defined to contract the torsion and torsion vector,
\begin{align}
  T = \frac{1}{4} T^{\rho\mu\nu} T_{\rho\mu\nu} - \frac{1}{4} T^{\rho\mu\nu} T_{\mu\nu\rho}
  - \frac{1}{4} T^{\rho\mu\nu} T_{\nu\rho\mu} - T^\mu T_\mu.
\end{align}
The EoM of the torsion is the algebraic and the interaction term of spinor is rewritten by the four-fermion interaction substituting the torsion equation.
Thus the minimal model of the ECSK theory is equivalent to the GR with the spinor field and four fermion interaction.
To obtain the extended theory of gravity, we consider the modification of the Einstein-Hilbert action by the arbitrary function of the scalar curvature, $F(R)$, in the same sense of the metric formalism.

\subsection{Cartan $F(R)$ gravity and equivalent scalar-tensor theory}
The action of Cartan $F(R)$ gravity is defined by replacing the curvature scalar $R$ 
in Einstein-Cartan theory with a general function $F(R)$ as follows,
\begin{align}\label{eqs:action FR}
S=\int d^4xe\left( \frac{{M_{\rm Pl}}^2}{2}F(R)+\mathcal{L}_{\psi}\right),
\end{align}
where Planck mass defines $M_{\rm Pl}^2={{c^4}/{8\pi G}}$.
A volume element is given by the determinant of the vierbein, $e$.
The degree of freedom (DoF) due to $F(R)$ modification increases from two in GR to three~\cite{Gong:2017bru,Myung:2016zdl,Moretti:2019yhs}.
While in Cartan $F(R)$ gravity, 
The increased DoF by the modification is encapsulated in the torsion.

To remove the DoFs that originated the torsion, we assume that the torsion is on-shell.
The Cartan equation of the Cartan $F(R)$ gravity is
\begin{align}
  \label{eqs:cartan eq in FR}
  M_\text{Pl}^2 \big[
    ({T^\nu}_{kl} - {e_l}^\nu T_k + {e_k}^\nu T_l) F'(R)
    + ({e_k}^\alpha{e_l}^\nu - {e_k}^\nu {e_l}^\alpha) \partial_{\alpha}F'(R)
  \big]
  + {S^\nu}_{kl}
  = 0,
\end{align}
where $F'(R) = \partial F(R) / \partial R$ and $S^\nu_{kl}$ is the spin density tensor which is given by
\begin{align}
  {S^\mu}_{kl}
  \equiv -\frac{\delta\mathcal{L}_\psi}{e\delta{\omega^{kl}}_\mu}
  = \frac{i}{2} \bar{\psi} {\gamma^{i} \gamma^{j} \gamma^{k} \psi = \frac{1}{2} \bar{\psi} \epsilon^{i j k l} \gamma^{5} \gamma_{l}} \psi.
\end{align}
From \eqref{eqs:cartan eq in FR} we can deduce that the torsion tensor is represented by the spin density tensor, the derivative of $F(R)$, and the vierbein
\cite{Montesinos:2020pxv,Inagaki:2022blm};
\begin{align}
  \label{eqs:torsion of FR}
  {T_{i j}}^{k}
  = \frac{1}{2} (\delta^{k}_{j} {e_{i}}^{\lambda} - \delta^{k}_{i} {e_{j}}^{\lambda})
    \partial_{\lambda} \ln F^{\prime}(R)
  - \frac{{{S}^{k}}_{i j}}{F^{\prime}(R){M_\text{Pl}}}.
\end{align}
We can extract the $F(R)$ correction and spinor field in the curvature scalar by Eq.~\eqref{eqs:Ricci scalar into non and torsion2},
\begin{align}
  \label{eqs::Cavature scalar::scalar spinor}
  R = R_{E}
    - \frac{3}{2} \partial_{\lambda} \ln F^{\prime}(R) \partial^{\lambda} \ln F^{\prime}(R)
    - 3 \nabla_E^{2} \ln F^{\prime}(R)
    - \frac{1}{4 F^{\prime}(R)^{2}{M_\text{Pl}}^2} {S}^{\mu \nu \rho} {S}_{\mu \nu \rho}.
\end{align}

Now we consider a certain class of Cartan $F(R)$ gravity expressed as $F(R)=R+f(R)$,
and define the scalar field, $\phi$, as
\begin{align}\label{eqs:def scalar field}
    \phi \equiv -\sqrt{\frac{3}{2}}{M_{\rm Pl}} \ln F'(R).
\end{align}
Insert it into Eq.\eqref{eqs:action FR} with Eq.\eqref{eqs::Cavature scalar::scalar spinor}, the gravity part of the action is rewritten
to be the Einstein-Hilbert term and the scalar field,
\begin{align}\label{eqs:action for R and scalar}
S 
= \int d^4x e\left(\frac{{M_{\rm Pl}}^2}{2}R_E-\frac{1}{2}\partial_{\lambda}\phi \partial^{\lambda}\phi-V(\phi)
+ \mathcal{L}_\psi
\right).
\end{align}
We assume that $\ln F'(R)$ vanishes at a distance,
consequently, a total derivative in \eqref{eqs::Cavature scalar::scalar spinor}, is omitted.
The potential is represented as a function of the scalaron field $\phi$ with Eq.(\ref{eqs:def scalar field}) as,
\begin{align} \label{eqs:potential for phi in general}
V(\phi) \equiv -\frac{{M_{\rm Pl}}^2}{2} f(R) \Big{|}_{R=R(\phi)}.
\end{align}

Thus, the equivalent scalar-tensor theory (\ref{eqs:action for R and scalar}) is derived without any conformal transformations.
In $F(R)$ gravity in metric formalism, the scalar field representation can be obtained by the conformal transformation, ${g}_{\mu\nu}=e^{-\phi}\tilde{g}_{\mu\nu}$ that extracts scalar degrees of freedom from the metric. 
On the other hand, the Cartan $F(R)$ gravity theory can introduce a scalar field in a way different from the conformal transformation
since the one additional DoF remains from 
substituting the Cartan equation into the Cartan $F(R)$ gravity.

The potential, $V(\phi)$, has a simpler representation from the one in the scalar-tensor theory obtained from conventional $F(R)$ gravity after the conformal transformation~\cite{Nojiri:2017ncd}.
Various potentials can be obtained from the function $f(R)$ in Cartan $F(R)$ gravity.
As an example, $f(R) = - \gamma R^2$ in Cartan $F(R)$ gravity derives a potential equal to the Starobinsky model~\cite{Inagaki:2022blm}.

In this paper, we focus on the logarithmic model~\cite{Inagaki:2023mxv} defined by
\begin{align}
    f(R) = -\alpha R \ln \Big( 1 + \frac{R}{R_0} \Big),
    \label{Model::Logarithmic}
\end{align}
with two parameters, $\alpha$ and $R_0$.
This model has the following two features.
As the coupling $\alpha$ increases, the potential obtained from Eq.~\eqref{Model::Logarithmic} approaches the Starobinsky model~\cite{Inagaki:2023dzn}.
For a small coupling, the model is approximated as $f(R) \sim - \frac{R_0}{e} (\frac{R}{R_0})^{1+\alpha}$ for a positive $\phi$ and the potential is given by
\begin{align}
    \label{eqs:potential_positive}
    V(\phi) = \frac{M_\text{Pl}^2 R_0}{2e} \Big\{
        \frac{e}{1+\alpha} \Big(1 - e^{-\sqrt{2/3} \phi/M_\text{Pl}}\Big)
    \Big\}^{\frac{1+\alpha}{\alpha}}.
\end{align}

In Ref.~\cite{Inagaki:2023mxv}, the accelerated expansion in the inflation and dark energy era can be explained simultaneously in this model.
To obtain the suitable CMB observables we relate the parameter $R_0$ to the inflation,
\begin{align}\label{eqs::model parameter}
    R_0= 2\Lambda_{\rm Inf} e^{-\frac{1}{\alpha}},
\end{align}
where the inflationary scale is taken to be $\Lambda_{\rm Inf} \sim (\SI{e15}{GeV})^2$.
Using Eq.~\eqref{eqs::model parameter}, the potential for the negative region, $\phi<0$, is expressed as a linear potential,
\begin{align} 
\label{eqs::pot:linear}
V(\phi) = -\sqrt{\frac{2}{3}} \alpha \Lambda_\text{Inf} e^{-1/\alpha} M_\text{Pl} \phi.
\end{align}
Therefore, the potential energy is exponentially suppressed for a small $\alpha$ in the negative $\phi$ region.
The dark energy scale is obtained for $\alpha\sim0.0039$,
\begin{align}
    \sqrt{\frac{2}{3}}\alpha\Lambda_{\rm Inf}e^{-\frac{1}{\alpha}}
    \sim\frac{\Lambda_{\rm Inf}}{10^{114}}\sim \Lambda_{\rm DE},
\end{align}
From Eq.~\eqref{eqs:potential_positive} and \eqref{eqs::pot:linear}, the scalar field in the positive region causes the inflation and then, being in the negative region, becomes dark energy in a quintessence scenario.
Therefore, we can consider quintessence inflation in the logarithmic model of $\alpha=0.0039$~\cite{Inagaki:2023mxv}.
For any $\alpha$, we can explore a unified scenario of inflation and neutron star physics. Thus, we adopt the potential in Eq.\eqref{eqs::pot:linear}.

\subsection{The correction of spinor field with the four-fermion interaction}
Not only the curvature scalar but the kinetic term of the spinor field also 
can be decomposed into the non-torsion part and the torsion parts,
\begin{align}\label{eqs:SC into non and torsion}
\omega_{ijk} ={\omega_\Diamond}_{ijk}+\frac{1}{2}\left(T_{ijk}+{T_{jki}}+{T_{kji}}\right).
\end{align}
The pure spin connection, $\omega_\Diamond$, can be expressed only in terms of viebein,
\begin{align*}
    {\omega_\Diamond}_{ijk}&=\frac{1}{2}(\Delta_{kij}-\Delta_{ijk}+\Delta_{jik}),
    \\
    \Delta_{kij}&=({e_i}^\mu{e_j}^\nu-{e_j}^\mu{e_i}^\nu)\partial_{\nu}e_{k\mu}.
\end{align*}
Substituting the \eqref{eqs::Cavature scalar::scalar spinor}, we can obtain the four-fermion interaction without lack of the Lorentz invariance.
The covariant derivative~\eqref{eqs::cov dev spinor} also holds the covariance by replacing $\omega$ with $\omega_\Diamond$.

Then we can rewrite the action of Cartan $F(R)$ gravity with spinor as scalar-tensor theory by definition of the scalar field~\eqref{eqs:def scalar field},
\begin{align}\label{eqs::Cartan+NJL}
S=&\int d^4x e{\Big{(}}\frac{{M_{\rm Pl}}^2}{2}R_E-\frac{1}{2}\partial_{\lambda}\phi\partial^{\lambda}\phi-V(\phi)-i \bar{\psi} \frac{1}{2}\left[\Slash{D}_{\diamond}-\overleftarrow{\Slash{D}}_\diamond\right] \psi-\frac{\Omega(\phi)}{4} \left(\bar{\psi} \gamma^{5} \gamma_{l} \psi\right)^2{\Big{)}}
\end{align}
where $\Slash{D}_{\Diamond}$ is the covariant derivative for the spinor replacing the spin connection, $\omega$ with $\omega_\Diamond$ in Eq.~\eqref{eqs::cov dev spinor}.
The coefficient including the scalar field becomes 
\begin{align}\label{eqs::Scalar Omega}
\Omega(\phi)=\frac{3}{4{M_{\rm Pl}}^2}\left(2e^{\sqrt{\frac{2}{3}}\frac{\phi}{{M_{\rm Pl}}}}-e^{2\sqrt{\frac{2}{3}}\frac{\phi}{{M_{\rm Pl}}}}\right).
\end{align}
The Lagrangian~\eqref{eqs::Cartan+NJL} has the local Lorentz symmetry because the pure spin connection remains covariant for the local Lorentz transformation.  
To begin with, even if we put it in GR or ECSK theory, i.e. $F(R)=R$, the spinor field has a four-fermion interaction due to the torsion contribution.
This interaction is called spin-spin interaction or Dirac-Heisenberg-Ivanenko-Hehl-Datta four-fermion interaction~\cite{Hehl:1974cn,Kerlick:1975tr,Gasperini:1986mv,Hehl:1971qi,Boos:2016cey}.
Thus, in Cartan $F(R)$ gravity, we obtained a coupling between the scalar field and the four-fermion.
By studying this in detail, we might demonstrate the existence of the scalar field from Cartan $F(R)$ gravity.

\section{Effective potential of scalaron}
To investigate the effect of the scalar field, we consider the effective potential of the scalar field, which is obtained by integrating out the fermion field with the spin-spin interaction.
Before performing the integration out, we execute Fiertz transformation to the spin-spin interaction in Eq.~\eqref{eqs::Cartan+NJL} as follows,
\begin{align}
\left(\bar{\psi} \gamma^{5} \gamma_{l} \psi\right)^2=2(\bar{\psi}\psi)^2+2(\bar{\psi}i\gamma^5\psi)^2-(\bar{\psi}\gamma^\mu\psi)^2.
\end{align}
After integrating out the spinor $\psi$, these terms may acquire nonzero vacuum expectation values (VEVs). For simplicity, we ignore the last term by assuming its VEV vanishes.

Not only Eq.~\eqref{eqs::Scalar Omega} but also the four-fermion interaction could be introduced to describe the low energy scale phenomenology, such as BCS superconductivity~\cite{PhysRev.106.162,PhysRev.108.1175}, chiral symmetry breaking in QCD~\cite{PhysRev.122.345,PhysRev.124.246},
or from a theoretical perspective, scenarios beyond Standard model (e.g. \cite{PhysRevD.19.1277,doi:10.1142/2170,HARADA20031}).
Let $\lambda$ be a coupling constant of four-fermion interaction, defined as $\lambda = \lambda_0 + \Omega$, where $\lambda_0$ is another four-fermion coupling constant.
Then we have
\begin{align}
  \begin{aligned}[b]
  S
  = \int d^4x e
    \Big{(} &\frac{{M_\text{Pl}}^2}{2} R_E 
        - \frac{1}{2} \partial_{\mu} \phi \partial^{\mu} \phi
        - V(\phi)
       \\
        &-i \bar{\psi} \frac{1}{2} \left[ \Slash{D}_{\diamond} - \overleftarrow{\Slash{D}}_\diamond \right] \psi
        + \frac{\lambda}{2} \big[
          (\bar{\psi}\psi)^2 + (\bar{\psi} i\gamma^5 \psi)^2
        \big]
    \Big{)},
  \end{aligned}
  \label{eqs::NJL+Cartan+spin}
\end{align}
By introducing auxiliary fields and performing a shift to eliminate the four-fermion interaction,
we obtain an equivalent model with Yukawa term instead of the four-fermion interaction term.
The action including auxiliary fields is
\begin{align}
\begin{aligned}[b]
    S
  = \int d^4x e
    \Big{(} &\frac{{M_\text{Pl}}^2}{2} R_E 
        - \frac{1}{2} \partial_{\mu} \phi \partial^{\mu} \phi
        - V(\phi)
        \\
        &-i \bar{\psi} \frac{1}{2} \left[ \Slash{D}_{\diamond} - \overleftarrow{\Slash{D}}_\diamond \right] \psi
        + \Big(
      \frac{1}{2\lambda} (\sigma^2 + \pi^2) + \bar\psi (\sigma + i\gamma^5 \pi) \psi
    \Big)
    \Big{)}
\end{aligned}
\end{align}
We can set $\pi = 0$ without the loss of generality by the rotation and redefinition for $\sigma$ and $\pi$.
Integrating out of the fermion with the cut-off regularization,
we obtain the one-loop corrected potential~\cite{Inagaki:1997kz},
\begin{align}
  \begin{aligned}[b]
    \mathcal{V} 
    =& \frac{1}{2\lambda} \sigma^2
    - \frac{1}{16\pi^2}
    \Big\{
        2 \Lambda_\text{cut-off}^2 \sigma^2
        + \sigma^4
        \Big(
            \ln\frac{\sigma^2}{\Lambda_\text{cut-off}^2} - \frac{1}{2}
        \Big)
    \Big\}
    \\
    &- \frac{R_E}{16\pi^2} \frac{2}{3} \Big\{
        \ln \Big( \frac{\Lambda_\text{cut-off}^2}{\sigma^2} \Big) - 1
    \Big\}  \sigma^2.
  \end{aligned}
\end{align}

The phase structure is determined by the coupling $\lambda$ and the curvature $R_E$.
In the symmetric phase, where the VEV is zero, the chiral symmetry of the model is preserved. In contrast, a finite VEV of the auxiliary field $\sigma$ breaks the chiral symmetry, which is present in the original model.
In the case where chiral symmetry is broken by the quantum correction, the fermion mass is generated dynamically.
The dynamical fermion mass is evaluated to solve the gap equation
\begin{align}
  \frac{1}{\lambda} - \frac{M^2}{4\pi^2} \Big(
    \frac{\Lambda_\text{cut-off}^2}{v^2} - \ln \frac{\Lambda_\text{cut-off}^2}{v^2}
  \Big)
  + \frac{R_E}{24\pi^2} \Big(
    2 - \ln \frac{\Lambda_\text{cut-off}^2}{v^2}
  \Big)
  = 0
  \label{Eqs::GapEquation}
\end{align}
where $v$ is the VEV of $\sigma$.
It is worth mentioning that the vacuum structure of the NJL model is determined by the value of the coupling constant $\lambda$. 
If $\lambda < \lambda_\text{cr}$, where $\lambda_\text{cr}$ is the critical value,
the system remains in the symmetric phase in Minkowski space. 
In curved space with a small coupling, the broken phase can occur only in regions with negative curvature. 
Therefore, introducing a non-vanishing $\lambda_0$ is important to realize the broken phase in this system.

Let $m$ be the VEV of the spinor pair as $\ev{\bar{\psi}\psi} = v/\lambda = m^3$,
the effective action is given by
\begin{align}
     \label{eqs::Action::Effective}
    \begin{aligned}[b]
    S_\text{eff} &= \int d^4x \sqrt{-g} \Big\{
        \frac{M_\text{Pl}^2}{2} (1 + \Pi) R_E 
        -\frac{1}{2} \partial^\mu \phi \partial_\mu \phi
        \\
        &+ {M_\text{Pl}}^2 \Lambda_\text{Inf} \alpha e^{-1/\alpha} \Big(\frac{\phi}{{M_{\rm Pl}}}\Big) 
        - \frac{3{m}^6}{8{M_{\rm Pl}}^2} \Big\{2e^{\sqrt{\frac{2}{3}}\frac{\phi}{{M_{\rm Pl}}}} - e^{2\sqrt{\frac{2}{3}}\frac{\phi}{{M_{\rm Pl}}}}\Big)
        \Big\},
    \end{aligned}
\end{align}
where
\begin{align}
  \Pi = \frac{M^2}{12\pi^2 M_\text{pl}^2} \Big\{
          \ln \Big( \frac{\Lambda_\text{cut-off}^2}{M^2} \Big) - 1
        \Big\}.
\end{align}
This term affects the Planck mass naively, however, we can absorb it by
the scale transformation, $g_{\mu\nu} \to (1 + \Pi) g_{\mu\nu}$ 
and field redefinition $\phi \to \phi/(1 + \Pi)^{1/2}$. Then we can
rewrite the potential as
\begin{align}
    \label{eqs::Potential::Total}
    U(\phi)
    = -U_1 \left(\frac{\phi}{M_{\rm Pl}}\right)
    + U_2 \left(2e^{\sqrt{\frac{2}{3}} (1 + \Pi)^{1/2} \frac{\phi}{M_\text{Pl}}}
    - e^{2\sqrt{\frac{2}{3}} (1 + \Pi)^{1/2} \frac{\phi}{M_\text{Pl}}}
    \right),
\end{align}
where the coupling constants are
\begin{align}
    U_1 = M_\text{pl}^2 \Lambda_\text{Inf} \alpha e^{-1/\alpha} (1 + \Pi)^{-3/2},
    &&
    U_2 = \frac{3m^6}{8 M_\text{pl}^2} (1 + \Pi)^{-2}.
\end{align}
The minimum of the potential can be determined by solving $U' = 0$. If the region of interest is limited to $-\phi \gg 1$, the solution simplifies to
\begin{align}
\phi_{\rm min} = \sqrt{\frac{3}{2}}{M_{\rm Pl}} \ln \Big(
\sqrt{\frac{3}{8}} \frac{U_1}{U_2(1 + \Pi)^{1/2}} \Big).
\end{align}
The correction in weakly curved space reduces the potential value; however, the effect is presumed to be very small. 
For example, at the QCD scale, $M \sim \SI{250}{MeV}$ and $\Lambda_\text{cut-off} \sim \SI{1}{GeV}$ yields $\Pi \sim 3.4 \times 10^{-41}$.
Thus, to the minimum value exists in the negative region, the parameters satisfy $U_2\gtrsim U_1$.

\section{
Modified TOV equation in Cartan $F(R)$ gravity
}
To explore the phenomena including the corrections from Cartan $F(R)$ gravity theory, we will analyze the changes in the properties of neutron stars.
%
The field equation corresponding to~\eqref{eqs::Action::Effective} can be
obtained by the variational principle of the metric $g_{\mu\nu}$ and scalar field $\phi$.
Taking account of a perfect fluid and a static and spherically symmetric metric,
\begin{align}\label{eqs::spherically sym}
    d^2s = -\varphi(r)dt^2 + h(r)^{-1} d^2r + r^2 (d^2\theta + \sin^2 \theta d^2\vartheta),
\end{align}
we have the field equation~\cite{Kase:2019dqc},
\begin{align}
    \label{eqs::eq for f(r)}
    & \frac{\varphi'}{\varphi} = -\frac{8\pi G}{h(r) r} \left[\frac{h(r)-1}{8\pi G}-r^2(P + P_s) \right],
    \\
    \label{eqs::eq of continuity}
    & \frac{dP}{dr}+\frac{\varphi'}{2\varphi}(\rho + P) = 0,
    \\
    \label{eqs::eq for mass}
    & \frac{dM}{dr}=4\pi (\rho + \rho_s) r^2,
    \\
    \label{eqs::EoM::scalar}
    & \frac{d^2\phi}{d^2r}+\left(\frac{2}{r}+\frac{1}{2}\frac{d}{dr}\ln(\varphi(r)h(r))\right)\frac{d\phi}{dr}-h(r)^{-1} U_{\phi}=0,
\end{align}
where $\rho$ and $P$ are the pressure and density of the matter, 
$\rho_s$ and $P_s$ are the pressure and density of the scalar field which are given by
\begin{align}
    \rho_s &= \frac{h(r)}{16\pi{G}} \left(\frac{d\phi}{dr}\right)^2 + U(\phi),
    \\
    P_s &= \frac{h(r)}{16\pi{G}} \left(\frac{d\phi}{dr}\right)^2 - U(\phi),
\end{align}
$U_{\phi}$ is the derivative for potential and
\begin{align}
    h(r) = 1 - \frac{2GM(r)}{r},
\end{align}

Eqs.~\eqref{eqs::eq for f(r)} - \eqref{eqs::EoM::scalar} can be used to investigate the structure of neutron star interiors.
The field equations, \eqref{eqs::eq for f(r)} - \eqref{eqs::EoM::scalar},
are solved numerically with the boundary condition.
To work on the numerical computation, we make variables dimensionless using the normalization factor,
\begin{align}
    \label{def::NormalizationFactor}
    \tilde{P} \equiv P / \rho_0 c^2,
    \quad 
    \tilde{\rho} \equiv \rho / \rho_0, 
    \quad
    \tilde{M} \equiv M / (\rho_0 {r_0}^3), 
    \quad 
    s \equiv r/r_0,
    \quad
    \tilde{\phi} \equiv \phi / M_{\rm Pl}, 
    \quad
    \tilde{U} \equiv U / (\rho_0 c^2).
\end{align}
Here we introduce the density, $\rho_0$, and distance, $r_0$.
The density defines $\rho_0=m_n n_0=\SI{1.6749e{14}}{g\cdot cm^{-3}}$, where $m_n$ is the neutron mass, $m_n=\SI{1.67e{-24}}{g}$, and the typical density of neutron stars, $n_0=0.1(\si{fm})^{-3}$.
The distance also defines $r_0=c/\sqrt{G\rho_0}=\SI{89.664}{km}$. 
The gravitational constant is also dimensionless by the density and the distance,
\begin{align}
    \label{def::NormalizationFactor::NewtonConstant}
    \tilde{G} \equiv G \rho_0 r_0^2 / c^2 = 0.996861.
\end{align}
For our investigation we have considered
The realistic equation of state (EoS) as well known as SLy~\cite{Haensel:2004nu} and APR~\cite{Potekhin:2017ufy} in a neutron star.
The EoS inside the neuron star can be represented by
\begin{align}
    \label{eqs::EoS::parameter}
    \begin{aligned}[b]
        \zeta(\xi)= & \frac{a_1+a_2 \xi+a_3 \xi^3}{1+a_4 \xi} f_0\left(a_5\left(\xi-a_6\right)\right)+\left(a_7+a_8 \xi\right) f_0\left(a_9\left(a_{10}-\xi\right)\right)
        \\
        &+\left(a_{11}+a_{12} \xi\right) f_0\left(a_{13}\left(a_{14}-\xi\right)\right)  +\left(a_{15}+a_{16} \xi\right) f_0\left(a_{17}\left(a_{18}-\xi\right)\right),    
    \end{aligned}
\end{align}
where $f_0(x)=(e^x+1)^{-1}$, and $\zeta, \xi$ are pressure and density defined by
\begin{align*}
    \zeta&=\log_{10} P/{\rm dyn\cdot cm^{-2}}=\frac{\ln \rho_0c^2/{\rm dyn\cdot cm^{-2}}}{\ln 10}+\frac{\ln \tilde{P}}{\ln 10},
\\
\xi&=\log_{10} \rho/ {\rm g\cdot cm^{-3}}=\frac{\ln \rho_0/{\rm g\cdot cm^{-3}}}{\ln 10}+\frac{\ln \tilde{\rho}}{\ln 10}.
\end{align*}
The parameters $a_1$ to $a_{18}$ in Eq. (38) are determined based on the corresponding equation of state (EoS).
Using dimensionless variables with \eqref{def::NormalizationFactor}, \eqref{def::NormalizationFactor::NewtonConstant} and EoS \eqref{eqs::EoS::parameter},
we have
\begin{align}
    \label{eqs::Eq::f}
    & \frac{\varphi'}{\varphi} = -\frac{8\pi \tilde{G}}{h(s)s} \left[ \frac{h(s)-1}{8\pi \tilde{G}}-s^2(\tilde{P}+\tilde{P}_s)\right],
    \\
    \label{eqs::Eq::rho}
    & \tilde{\rho}' + \frac{\tilde{\rho}}{2\tilde{P}}\left(\frac{d\zeta}{d\xi}\right)^{-1}\frac{\varphi'}{\varphi}(\tilde{\rho}+\tilde{P}) 
    = 0,
    \\
    \label{eqs::Eq::phi}
    & \tilde{\phi}''+\left(\frac{2}{s}+\frac{1}{2}\frac{d}{ds}\ln(\varphi(s)h(s))\right)\tilde{\phi}'-h(s)^{-1}8\pi\tilde{G}\tilde{U}_{\tilde{\phi}}
    = 0,
    \\
    \label{eqs::Eq::Mass}
    & \tilde{M}' = 4\pi(\tilde{\rho}+\tilde{\rho}_s) s^2.
\end{align}
The dimensionless expression of the potential of Eq.\eqref{eqs::Potential::Total} becomes
\begin{align}\label{eqs::potential::total::dimless}
    \tilde{U}(\phi) = \frac{U(\phi)}{\rho_0c^2}=-\tilde{U}_1\left(\frac{\phi}{{M_{\rm Pl}}}\right)+\tilde{U}_2\left(2e^{\sqrt{\frac{2}{3}}\frac{\phi}{{M_{\rm Pl}}}}-e^{2\sqrt{\frac{2}{3}}\frac{\phi}{{M_{\rm Pl}}}}\right).
\end{align}
Each parameter, $\tilde{U}_1$ and $\tilde{U}_2$ can be evaluated by the constant, $\Lambda$ from the logarithmic model in Cartan $F(R)$ gravity and the fermion condensate,$\bar{\psi}\psi$ from NJL model,
\begin{align}
    &\tilde{U}_1=\left(\frac{{M_{\rm Pl}}^2\Lambda}{\rho_0c^2}\right)\sim8.2\times10^{21}\left(\frac{\Lambda^{1/2}}{\SI{1}{eV}}\right)^2\sim1.38\times10^{33}\left(\frac{U_1}{\SI{1}{eV}}\right)^4,
    \\
    &\tilde{U}_2=\frac{3{m}^6}{8{M_{\rm Pl}}^2\rho_0c^2}\sim2.3\times10^{-40}\left(\frac{\ev{\bar{\psi}\psi}}{(\SI{100}{MeV})^3}\right)^2.
\end{align}
\begin{figure}
    \centering
    \includegraphics[width=0.6\linewidth]{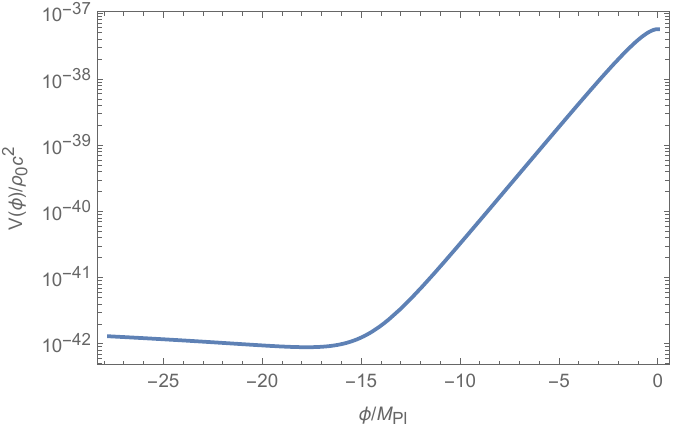}
    \caption{The potential, $U(\phi)=-4.7\times10^{-44}\phi+
    5.7\times10^{-38}\left(2e^{\sqrt{\frac{2}{3}}{\phi}}-e^{2\sqrt{\frac{2}{3}}{\phi}}\right)$. }
   \label{Fig::potential}
\end{figure}

As a concrete example, we apply the most typical value in our universe.
The energy scale of linear potential in Eq.~\eqref{eqs::potential::total::dimless} is
the dark energy scale, $\Lambda_\text{DE}\sim(\SI{e-33}{eV})^2$. 
Then the parameter value can be calculated as $\tilde{U}_1=4.7\times10^{-44}$.
We also consider the VEV for the spinor as the scale of chiral symmetry breaking in QCD, $\ev{\bar{q}q}\sim(\SI{250}{MeV})^3$. 
Then the other parameter becomes $\tilde{U}_2\sim5.7\times10^{-38}$.

We plot the potential with dark energy and QCD scale in Figure~\ref{Fig::potential}.
We can find that the potential has a minimum value at $\phi_{\rm min} = -17.8$.

\section{Numerical results}
Following the below setup, we show the numerical results of the mass-radius relation of the neutron star.
We use the shooting method to solve eqs.~\eqref{eqs::Eq::f}-\eqref{eqs::Eq::Mass} because
the exterior region of compact star must match the Schwarzschild solution,
and the scalar field value at the center of neutron stars cannot be determined a priori.
However, at least at a point at infinity, the scalar field resides at the minimum of the potential.
Therefore, in our calculation, we assume that the scalar field takes a value close to the minimum, $\phi_\text{min} = \ln \big(\sqrt{\frac{3}{8}}\frac{U_1}{U_2} \big)$, at the stellar surface, $s_\text{sur}$.
For the numerical calculation of the scalar field, we use the shooting method with this condition instead of the ODE solver with the initial condition.
For the other condition of variables, we assume $\rho(s=10^{-6})=\rho_c$, $\varphi(s=10^{-6})=1$, $M(s=10^{-6})=4\pi/3(10^{-6})^{3}\rho_c$ and $\phi'(s=10^{-6})=0$.
\begin{figure}
    \centering
    \includegraphics[width=1\linewidth]{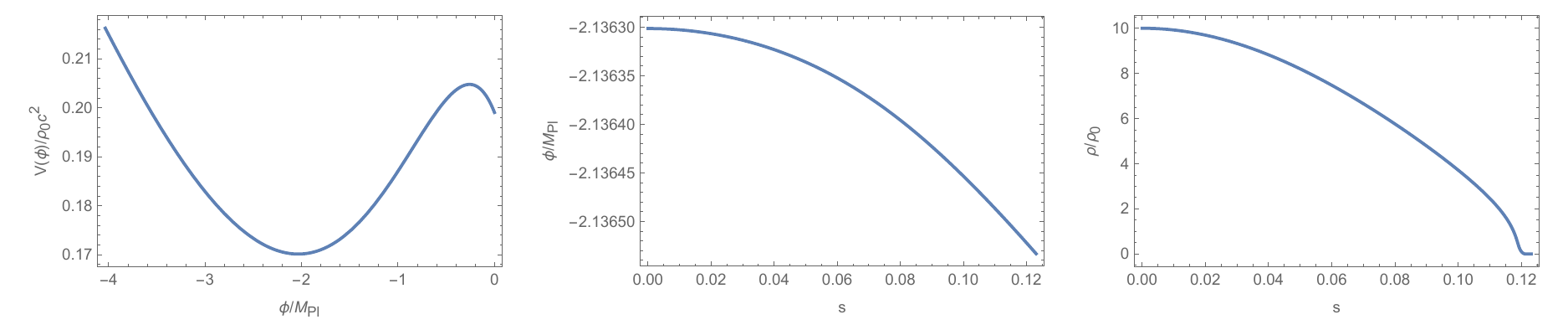}
    \caption{The potential, $U(\phi)=-5\times10^{-2}\phi+5\times10^{-1.4}\left(2e^{\sqrt{\frac{2}{3}}{\phi}}-e^{2\sqrt{\frac{2}{3}}{\phi}}\right)$(Left). The scalar field value inside the neutron star(Middle). The density of the neutron star(Right).
    The central density is $\rho(10^{-6})=10\rho_0\sim2\times10^{15} {\rm g\cdot cm^{-3}}$and $\phi_{\rm sur}=1.05\phi_{\rm min}=-2.137$.}
   \label{Fig::potential+scalar}
\end{figure}

Let us consider the case when the potential parameters are $\tilde{U}_1=5\times  10^{-2}$ and $\tilde{U}_2=5\times10^{-1.4}$.
To determine the surface of the star, the density at the surface is needed to stipulate in advance.
The density at the outer crust of the neutron star is estimated $\rho\sim 10^{11}-10^{4}~\si{g\cdot cm^{-3}}$~\cite{Haensel:1993zw,Chamel:2008ca,Anil:2020lch}.
The outer crust has a wide range of density, but the rate of decrease is also large, so the effect of the difference of the choice of density at the surface on the radius of the star seems to be small.
Therefore, the surface density is set to $\rho_{\rm sur}=\SI{e10}{g\cdot cm^{-3}}$ in this paper, considering the efficiency of the calculation.

As an example, Figure~\ref{Fig::potential+scalar} shows the potential and the calculation results of the shooting method when the central density is $\rho_{c}=10\rho_0\sim \SI{2e15}{g\cdot cm^{-3}}$.
From the stationary condition, we lead that the minimum value is $\phi_{\rm min}=-2.035$.
The scalar field value at the surface has a 5$\%$ fluctuation $\phi_{\rm sur}=1.05\phi_{\rm min}\sim-2.137$.
Then the surface is estimated at $s_{\rm sur}\sim0.1231$, thus $R\sim\SI{11.04}{km}$.
We obtain the result that the total mass is $2.025M_{\odot}$ by integrating the density to the surface.
$M_{\odot}=\SI{1.9884e{33}}{g}$ is the solar mass.

Next, we assess the effect of fluctuations in the value of the scalar field at the surface.
The mass-radius relation is drawn by computation with the several values of the central density.
The range of the density is set to $\rho_{c}/\rho_0=10^{0.3}-10^{1.5}$.
The degree of fluctuation of the scalar field values at the surface is expressed using $\Delta$ defined by
\begin{align*}
    \Delta\equiv \frac{\phi_{\rm surface}}{\phi_{\rm min}}-1.
\end{align*}

\begin{figure}
    \centering
    \includegraphics[width=0.6\linewidth]{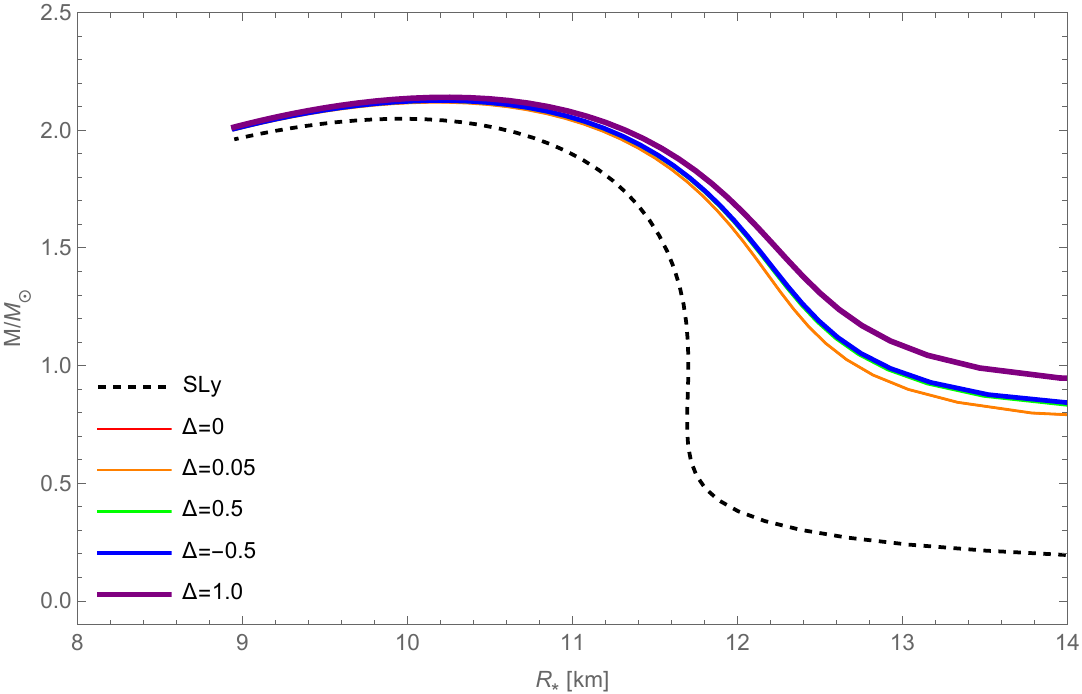}
    \caption{Mass-radius relation with $\tilde{U}_1=5\times10^{-2},\tilde{U}_2=5\times10^{-1.4}$. The difference of the scalar field value on the surface, $\Delta=0, 0.05, 0.5, -0.5, 1.0$.}
   \label{Fig::Mass-Radius::scalar::SLY::fluctuation}
\end{figure}
Figure~\ref{Fig::Mass-Radius::scalar::SLY::fluctuation} shows the mass-radius relation of a neutron star in the presence of a scalar field with $\Delta=0, 0.05, 0.5, -0.5, 1.0$ fluctuations at the surface from the minimum potential point.
Even when $\Delta=1$, i.e., differing by a factor of two from the minimum, the mass of the star with the larger radius is only about $0.2M_{\odot}$ bigger to the smaller $\Delta$.
This result indicates that differences in scalar field fluctuations do not significantly affect the properties of neutron stars.
Therefore, it is clear that the energy scale of the potential of the scalar field is important for the effect on neutron stars.

\begin{figure}
    \centering
    \includegraphics[width=0.6\linewidth]{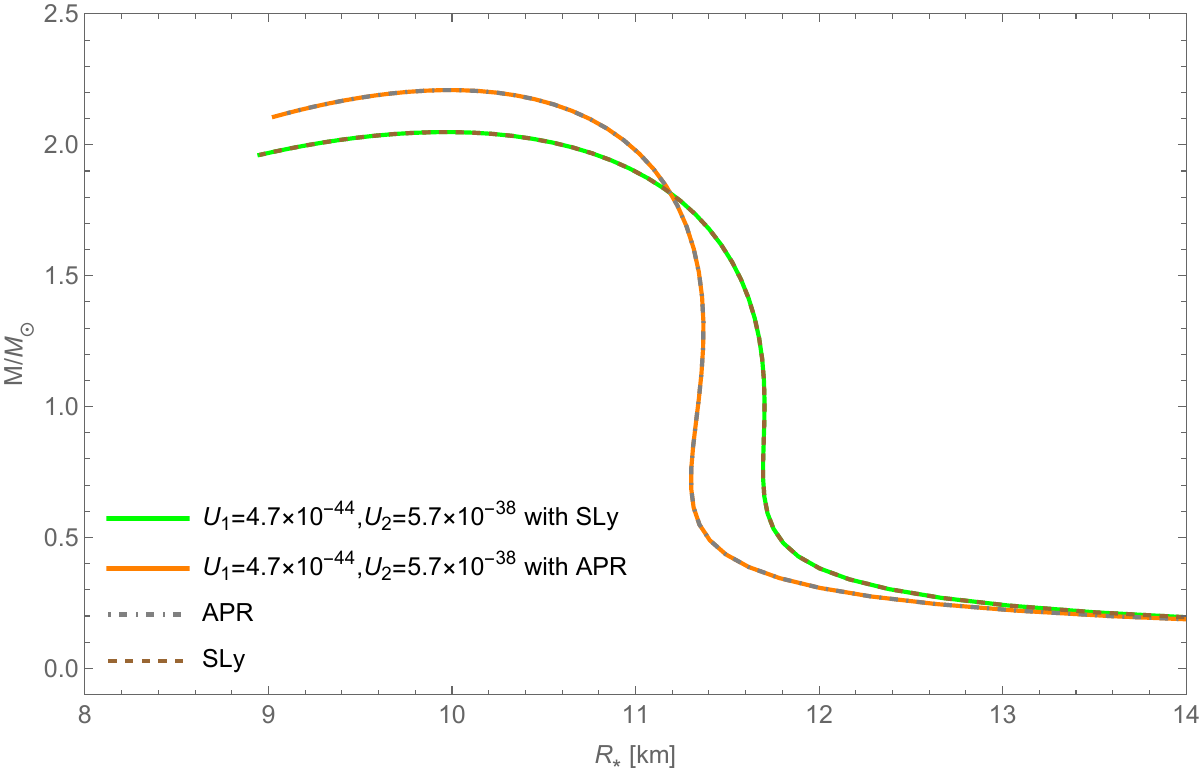}
    \caption{Mass-radius relation. The dashed line shows the EoSs, SLy, and APR results. The Orange and Green line shows the results by the potential~\eqref{eqs::potential::total::dimless} with the dark energy and chiral symmetry breaking scale.}
   \label{Fig::Mass-Radius::DEQCD::SLYAPR}
\end{figure}

In Figure~\ref{Fig::Mass-Radius::DEQCD::SLYAPR} we evaluate the properties of the neutron star with the dark energy and chiral symmetry breaking scale for the parameters of the potential.
The initial value of the scalar field is set to $\phi_{\rm sur}=1.05\phi_{\rm min}$.
On the dark energy scale, the correction to neutron stars due to the scalar field is quite small.
Therefore, it is difficult to confirm by observation of neutron stars in the case where the dark energy due to quintessence is realized in the logarithmic model of Cartan $F(R)$ gravity theory.

\begin{figure}
    \begin{minipage}{0.48\linewidth}
    \centering 
    \includegraphics[,scale=0.36]{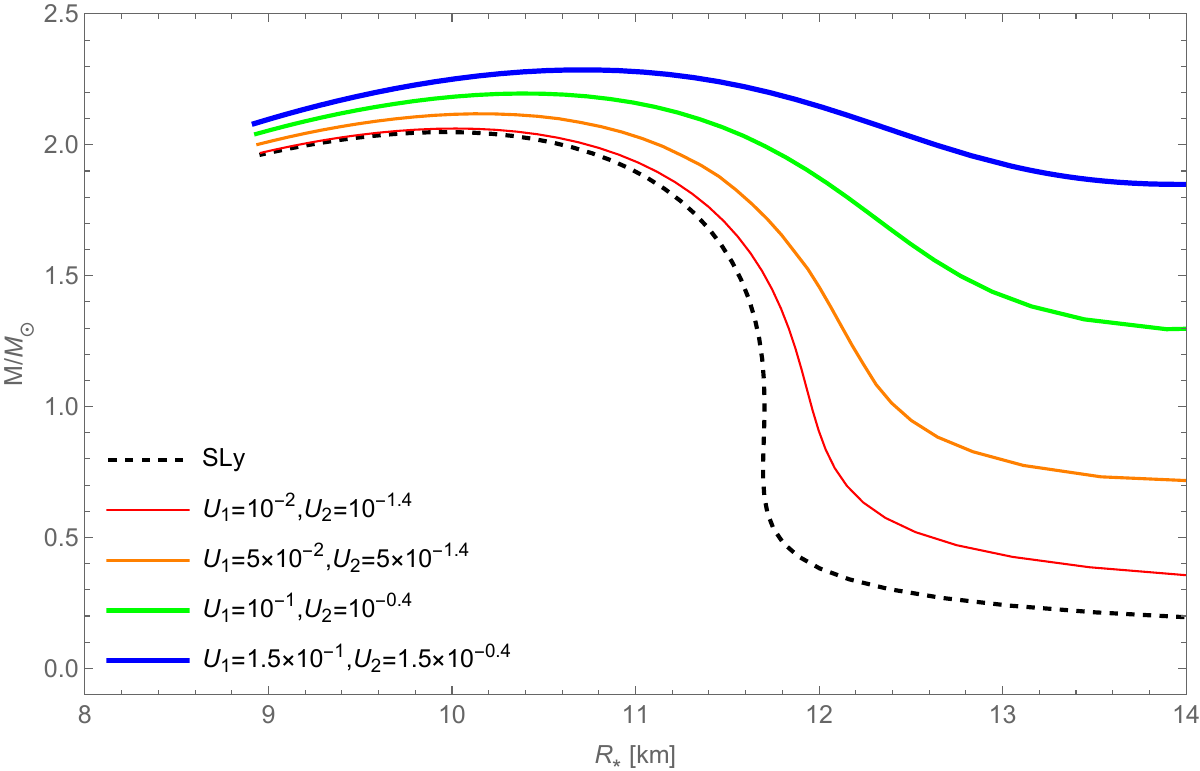}
    \subcaption{SLy}
    \end{minipage}
    \begin{minipage}{0.48\linewidth}
    \centering
    \includegraphics[scale=0.36]{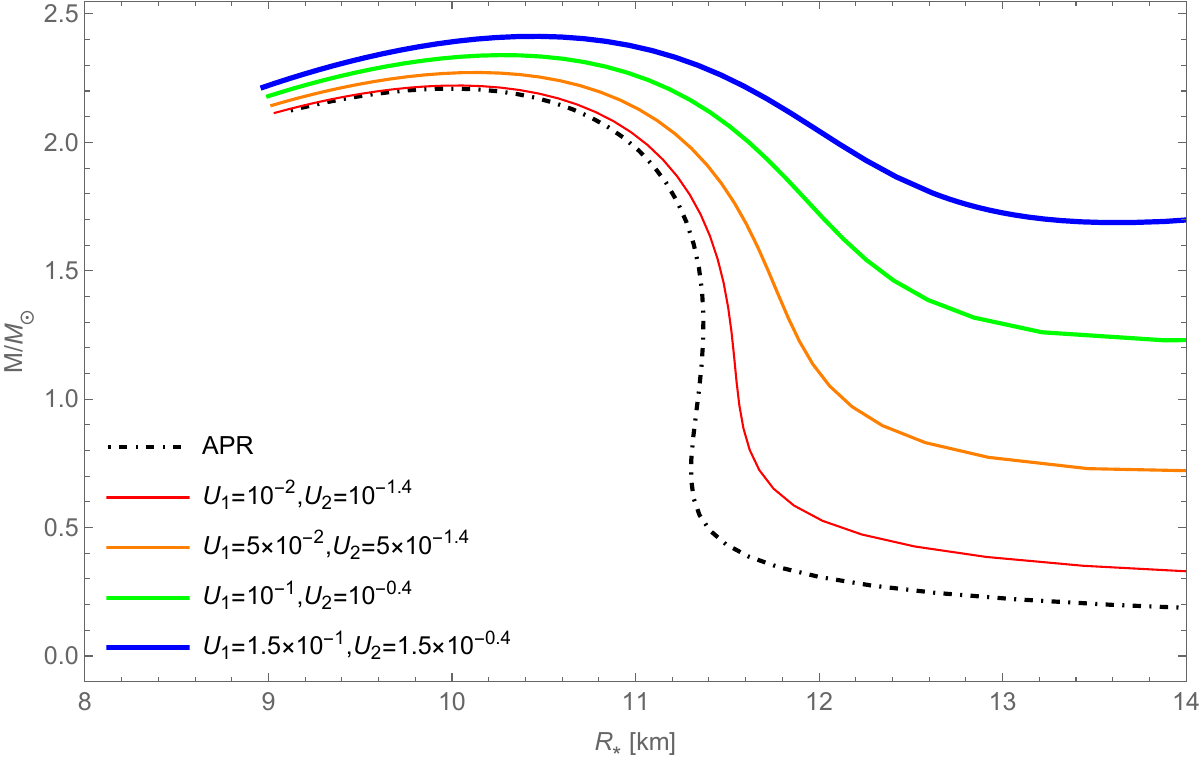}
    \subcaption{APR}
    \end{minipage}
    \caption{Mass-Radius relation for the potential of the scalar field in Eq.~\eqref{eqs::potential::total::dimless}. (a)SLy and (b)APR are used for the EoS of the neutron star. The parameters of potential are  $\tilde{U}_1=10^{-2},5.0\times10^{-2},10^{-1},1.5\times10^{-1}$ and $\tilde{U}_2=10^{-1.4},5.0\times10^{-1.4},10^{-0.4},1.5\times10^{-0.4}$.}  
    \label{Fig::Mass-Radius::scalar}
\end{figure}

\begin{figure}
    \begin{minipage}{0.48\linewidth}
    \centering 
    \includegraphics[,scale=0.4]{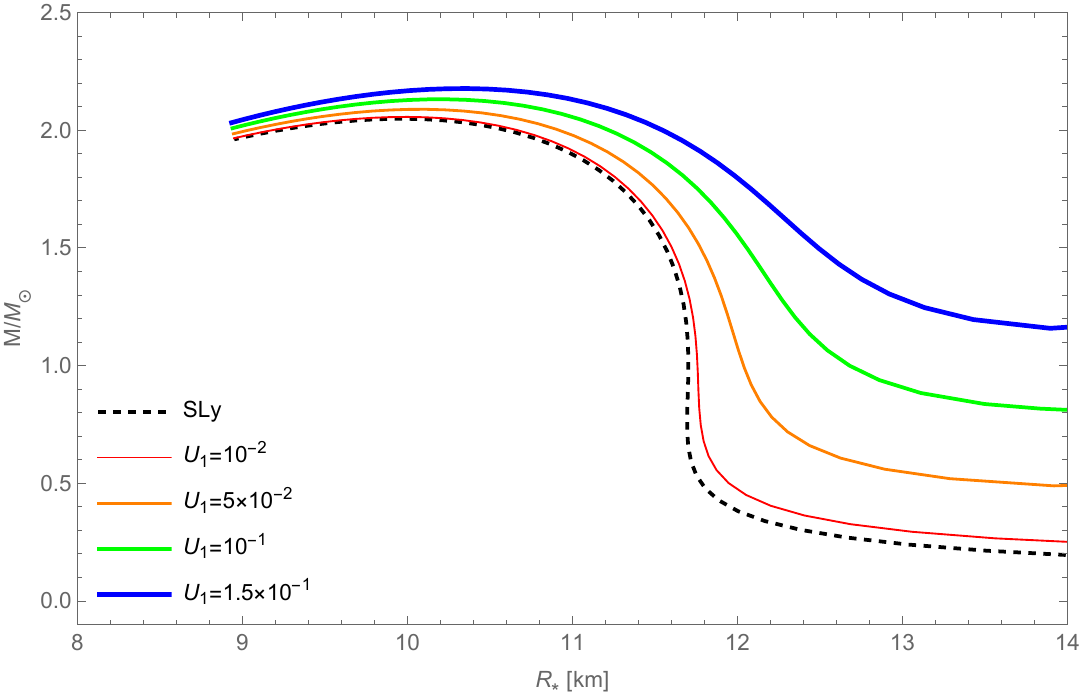}
    \subcaption{SLy}
    \end{minipage}
    \begin{minipage}{0.48\linewidth}
    \centering
    \includegraphics[scale=0.4]{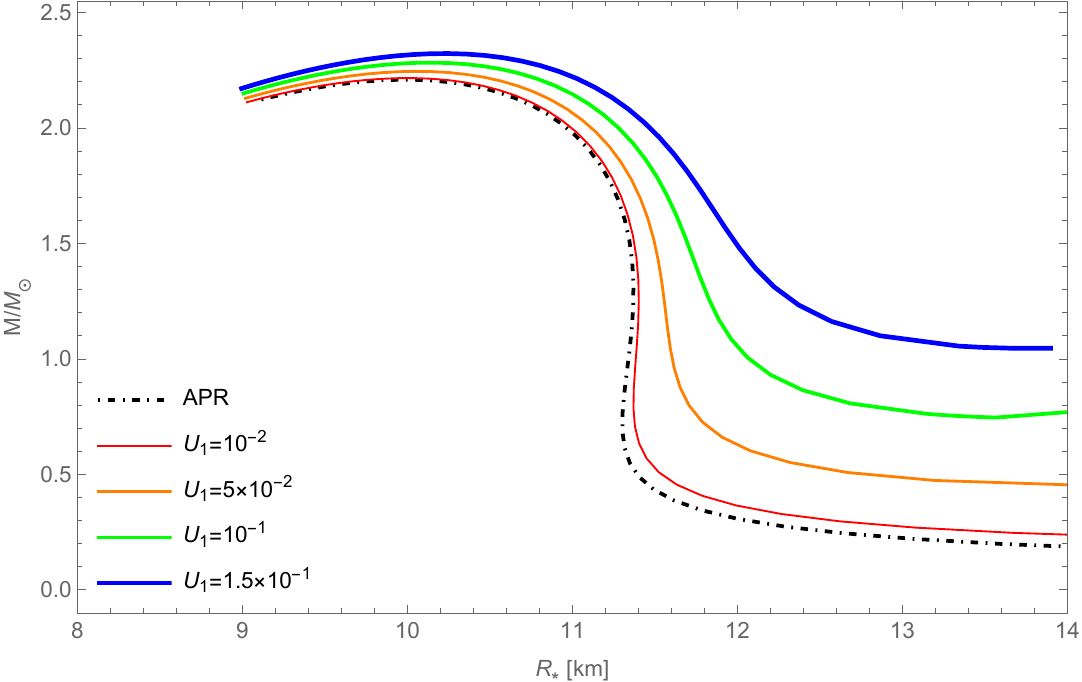}
    \subcaption{APR}
    \end{minipage}
    \caption{Mass-Radius relation for the linear potential of the scalar field. (a)SLy and (b)APR are used for the EoS of the neutron star. The parameters of potential are  $\tilde{U}_1=10^{-2},5.0\times10^{-2},10^{-1},1.5\times10^{-1}$.} 
    \label{Fig::Mass-Radius::scalar::Linear}
\end{figure}

Next, we investigate cases in which the contribution from Cartan $F(R)$ gravity is significant.
The main contribution in potential is parameter $\tilde{U}_1$. Therefore, we numerically analyze the neutron star for several values of $\tilde{U}_1$.
For the parameter, $\tilde{U}_2$, we tune that the potential has the minimum value of the scalar field at $\phi_{\rm min}\sim -2$.
Figure~\ref{Fig::Mass-Radius::scalar} shows the mass-radius relation for SLy and APR.
In both cases, large corrections from the scalar field are obtained, especially for neutron star masses with a radius greater than 12km.
It can be seen to have a significant impact, especially with respect to theoretical limits on the minimum mass of neutron stars.
Figure~\ref{Fig::Mass-Radius::scalar::Linear} shows the mass-radius relation in the case of the fermion has no dynamical mass.
The scalar field only has the potential introduced by the logarithmic model~\eqref{eqs::pot:linear}.
For comparison with Figure~\ref{Fig::Mass-Radius::scalar}, we assume that the boundary condition for the scalar field is $\phi=-2$.
It can be seen that the Mass-Radius curve shifts to reduce to the mass in Figure~\ref{Fig::Mass-Radius::scalar::Linear} than in Figure~\ref{Fig::Mass-Radius::scalar}.
The result is due to a decrease in the potential value due to the absence of chiral condensation-derived potential superposition.
The scalar field on only the linear potential would reach zero with time evolution and the effects of the scalar field disappear.
The presence of a potential due to chiral symmetry breaking is crucial for the effects on neutron stars to be observed.

The upper and lower limits for the mass of the neutron star at each parameter that can be read from figure~\ref{Fig::Mass-Radius::scalar} are summarized in Table~\ref{Table::Observables::numerical results}.
At the end of the table, we also include mass constraints from pulsar observations of binary stars~\cite{Lattimer:2012nd}.
Observations show that most neutron stars have $1.2-1.5M_{\odot}$ masses. Even the smallest masses are roughly equivalent to the mass of the Sun.
On the other hand, the lower limit of mass calculated by typical EoS is predicted to be $0.2M_{\odot}$ in the case of a large radius.
In the presence of a $\SI{78}{MeV}$ scale potential brought about by Cartan $F(R)$ gravity, the minimum mass of the neutron star rises to $0.7M_\odot$.
Conversely, for scales above $\SI{90}{MeV}$ 
the minimum mass exceeds the solar mass and is inconsistent with observations.
Therefore, for scales below $\SI{80}{MeV}$, the Cartan $F(R)$ gravity effect on neutron stars is consistent with observations.
In particular, for scales near $\SI{80}{MeV}$, the theoretical limits are close to observational constraints.

It should be noted that the parameter, $\tilde{U}_2$, must be greater than $\tilde{U}_1$ for the potential to have a minimum value from the discussion in Sec. 2.
Given that, the parameter must be $U_2>5\times10^{-2}$ and must have a value greater than $(\SI{250}{TeV})^3$ in the vacuum expectation value.
That is, in addition to Cartan $F(R)$ gravity modification, a mechanism of chiral condensation at the ${100-1000}~\si{TeV}$ scale is required.
If a neutron star with a large radius has a heavy mass, it might suggest a modification of gravity within the framework of Cartan $F(R)$ theory and indicate the existence of chiral symmetry breaking at the TeV scale.

\begin{table}[b]
    \centering
    \begin{tabular}{cccccc}
    \hline
         $\tilde{U}_1$ & $U_1^{1/4}$ [MeV] & EoS & $M_{max}$ [${M_{\odot}}$]  & $M_{min}$ [${M_{\odot}}$]  \\ \hline
      - & - & SLy &2.05 & 0.214 &  \\
         & ~ & APR & 2.21 & 0.200 &  \\
       $10^{-2}$ & $\SI{52}{}$ & SLy &2.06 & 0.355 &  \\
         & ~ & APR & 2.22 & 0.320 &  \\ 
         $5\times10^{-2}$ & $\SI{78}{}$ & SLy &  2.12& 0.712 &  \\ 
         & ~ & APR &2.27  & 0.713 &  \\ 
        $10^{-1}$  & $\SI{92}{}$ & SLy & 2.20 & 1.30 &  \\ 
         & ~ & APR &2.34  &1.22  &  \\ 
        $1.5\times10^{-1}$ & $\SI{102}{}$ & SLy & 2.29 & 1.83 &  \\ 
         & ~ & APR & 2.39 &1.69  &  \\ 
         \hline
         Constraints~\cite{Lattimer:2012nd} & - & - &2.7$\pm$0.2  &  1.0$\pm$0.1& \\ \hline
    \end{tabular}
    \caption{Theoretical limit of the maximum and minimum mass of the neutron star for each EoS.}
    \label{Table::Observables::numerical results}
\end{table}

\section{Conclusions}
We have investigated the mass-radius relation of the neutron star in Cartan $F(R)$ gravity.
Cartan $F(R)$ gravity theory derives an equivalent scalar-tensor theory by the modified Cartan equation.
It also leads the interaction between the scalar field and spinor as the four-fermion interaction term.

In the metric $F(R)$ theory of gravity, the scalar field is coupled to the matter field through a conformal transformation. 
In contrast, in Cartan $F(R)$ gravity theory, the interaction can be restricted due to its origin in local Lorentz symmetry. 
Thus, it becomes possible to distinguish the theoretical predictions.

We have used the effective potential~\eqref{eqs::Potential::Total} 
which is derived by the four-fermion interaction and spontaneous breaking of the chiral symmetry.
The TOV equation of the neutron with scalar field has been numerically analyzed.
We first addressed corrections to neutron stars by dark energy and QCD scales based on the Standard Model of cosmology and particle theory.
When using the dark energy parameter, it is found that the contribution of the scalar field is negligibly small (see Figure~\ref{Fig::Mass-Radius::DEQCD::SLYAPR}). 
As a result, it becomes difficult to observe any trace of modified gravity.
While, in Figure~\ref{Fig::Mass-Radius::scalar} and Table~\ref{Table::Observables::numerical results}, we see that the contribution of the scalar field increases the neutron star mass. Notably, for $U_1 \sim \SI{80}{MeV}$, the minimum mass of a neutron star is approximately one solar mass, which is consistent with observations, as no light neutron stars have been discovered.

We are interested in the scalar field as a candidate of dark matter.
The mass of the scalar field is found by the second-order derivative of the potential at the minimum point, $m^2_{\phi}=\left.U''\right|_{\phi_{\rm min}}\sim (U_1)^{1/2}$.
It is possible that a scalar field with a mass in the MeV scale range 
could become a part of the dark matter~\cite{Hooper:2007tu, Ho:2012ug, Choudhury:2019tss}.
We hope to report some results in future.

\acknowledgments
Authors would like to thank Tomohiro Inagaki,  Taishi Katsuragawa, and Muhammad Azzam Alwan for their valuable comments.


\bibliographystyle{JHEP}
\bibliography{ref}
\end{document}